\documentclass{article}
\usepackage{authblk}
\usepackage[a4paper, left=3cm, right=3cm, top=2.5cm, bottom=2.0cm]{geometry} 
\usepackage{mathptmx}   
\usepackage{helvet}          
\usepackage{courier}        

\usepackage{tabularx}
\newcolumntype{C}{>{\centering\arraybackslash}X}
\usepackage{caption}
\usepackage{amsmath}
\usepackage{amssymb}
\usepackage{textcomp}
\usepackage{amsfonts}
\usepackage{float}
\usepackage{wrapfig}
\usepackage{sidecap}
\usepackage{graphicx}
\usepackage{helvet}
\usepackage{color}
\usepackage[authoryear]{natbib}
\usepackage{textcomp}
\usepackage{siunitx}
\usepackage{upgreek}
\usepackage{sidecap}

\usepackage{lineno}
\usepackage{setspace}
\usepackage[latin1]{inputenc}
\nopagebreak
\title{Response and noise correlations to complex natural sounds\\ in the auditory midbrain}
\author[1,2,3]{Dominika Lyzwa \footnote{Corresponding author: dlyzwa@gwdg.de}}
\author[3,4]{Florentin W\"org\"otter}
\affil[1]{Max Planck Institute for Dynamics and Self-Organization, G\"ottingen, Germany}
\affil[2]{Institute for Nonlinear Dynamics, Physics Dep., Georg-August-University, G\"ottingen}
\affil[3]{Bernstein Focus Neurotechnology, G\"ottingen}
\affil[4]{Institute for Physics-Biophysics, Georg-August University, G\"ottingen}

\begin{document}

\maketitle



\textbf{\Large{Abstract}}\newline
\\
How natural communication sounds are spatially represented across the inferior colliculus, the main center of convergence for auditory information in the midbrain, is not known. The neural representation of the acoustic stimuli results from the interplay of locally differing input and the organization of spectral and temporal neural preferences that change gradually across the nucleus. This raises the question how similar the neural representation of the communication sounds is across these gradients of neural preferences, and whether it also changes gradually. Multi-unit cluster spike trains were recorded from guinea pigs presented with  a spectrotemporally rich set of eleven species-specific communication sounds. Using cross-correlation, we analyzed the response similarity of spiking activity across a broad frequency range for similarly and differently frequency-tuned neurons. Furthermore, we separated the contribution of the stimulus to the correlations to investigate whether similarity is only attributable to the stimulus, or, whether interactions exist between the multi-unit clusters that lead to correlations and whether these follow the same representation as the response similarity. We found that similarity of responses is dependent on the neurons' spatial distance for similarly and differently frequency-tuned neurons, and that similarity decreases gradually with spatial distance. Significant neural correlations exist, and contribute to the response similarity. Our findings suggest that for multi-unit clusters in the mammalian inferior colliculus, the gradual response similarity with spatial distance to natural complex sounds is shaped by neural interactions and the gradual organization of neural preferences.
\section{Introduction} 
\begin{figure}[h!,t,b,p]\centering\includegraphics[width=0.42\linewidth]{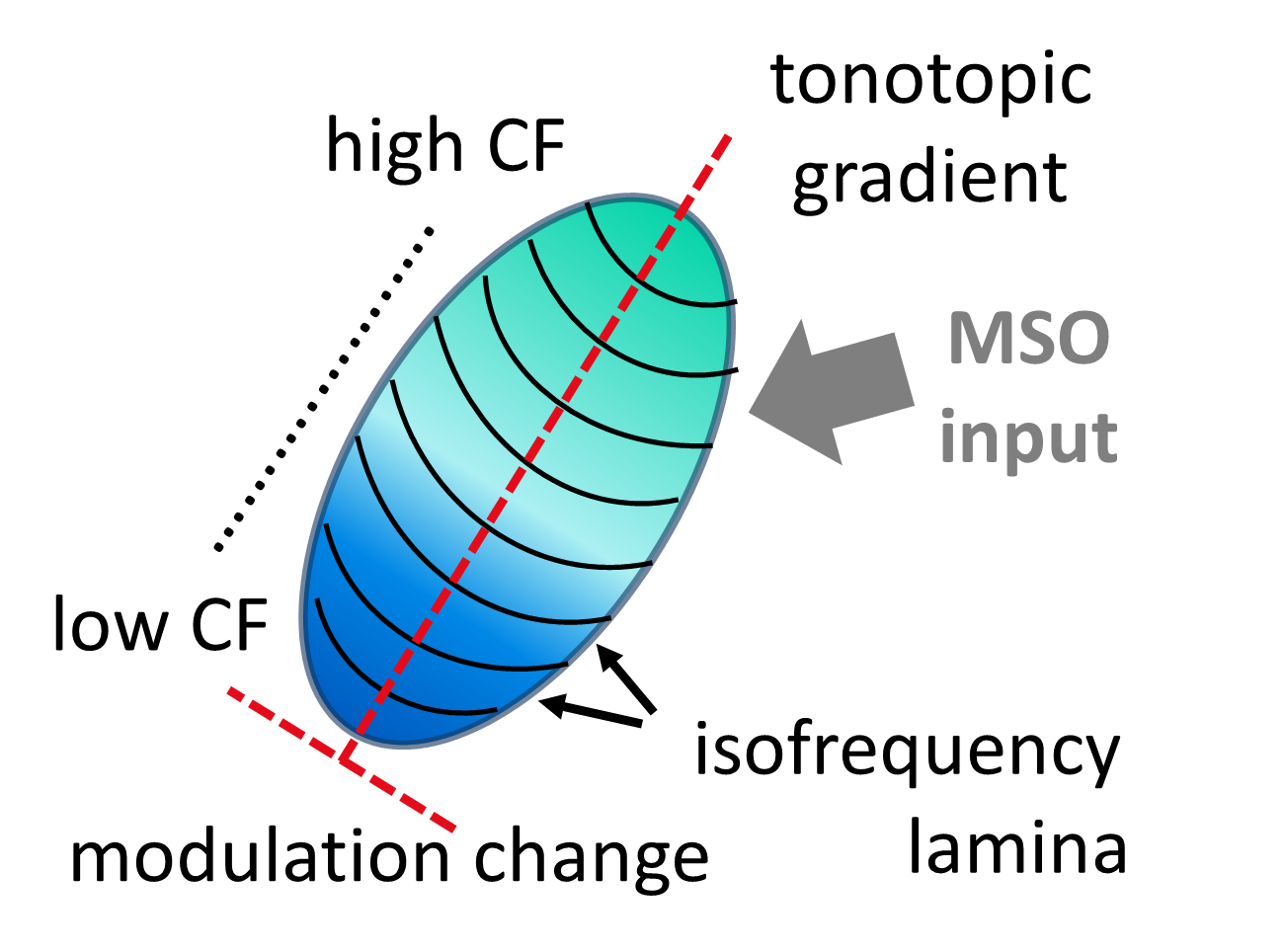}\caption[Schematic of IC.]{\textbf{Schematic of the central inferior colliculus displaying the tonotopic gradient and isofrequency laminae.} The inputs to the inferior colliculus are very diverse \citep{DougChapter}, e.g. the Medial Superior Olive (MSO) projects mainly to low and middle frequency regions.}\label{fig:ICschem}\end{figure}

A neuron's response is shaped by all the inputs it receives, as well as by the integration and processing of this input, hence by the neuron's stimulus preferences. The inferior colliculus is the main center of convergence in the auditory midbrain \citep{IrvineMainIC1992}. It receives and integrates diverse preprocessed inputs from essentially all ascending auditory brainstem nuclei \citep{AitkinICallconv1984, MalmiercaNotAllConvIC2002} that terminate on different locations within the central inferior colliculus (ICC) \citep{DougChapter}. Differences exist e.g. for high and low frequency regions as well as for caudal or rostral regions. Information about interaural time differences from the medial superior olive for example is mainly projected to low and middle frequency regions \citep{DougChapter}. Neural preferences to stimulus frequency and modulation are mainly organized gradually within the ICC \citep{Merzenich1974, LangnerPeriod1988b, Langner2002}. In the tonotopic gradient (Fig.\ref{fig:ICschem}) low frequencies are represented dorsol laterally and high frequencies ventral medially \citep{RoseTonotopy1961, Merzenich1974}. Along this tonotopic gradient, the stimulus frequency which elicits the highest spiking response gradually increases. For a given intensity this is called the best frequency (BF) and for the overall lowest spike-eliciting intensity this is the characteristic frequency (CF). Oriented approximately orthogonal to this frequency gradient are laminae that contain neurons with very similar best frequencies within a range of 1/3 octave, the isofrequency laminae \citep{LangnerLaminar1997}. Strong indications for a concentric gradient within laminae of preferred amplitude modulation frequencies for the sound envelope have been provided \citep{LangnerPeriod1988b, Langner2002, Rees2011}. The ICC has also been shown to be essential for extracting time-varying spectrotemporal information \citep{Monty2002} and therefore might be important for processing of complex sounds.\\

The neural response, the representation of acoustic stimuli in the ICC results from the interplay of the locally differing and heterogeneous input and the spatially gradual change of spectrotemporal neural preferences. Along these gradients the sound is filtered either for the same spectral content or for same amplitude modulations. Thus, the question arises how for complex, spectrotemporally rich sounds, such as speech or vocalizations, the neurons' output is organized across this main convergence center. Neural interactions of connected neurons could further contribute to the specific organization of the neural representation in this nucleus, hence the output of the neurons to vocalization stimuli. Comparing response correlations and neural correlations which result from connected neurons allows the evaluation of whether similarity of responses is shaped by the underlying neural structure rather than by the stimulus input to the neurons.\\ 
In this work we investigate the dependence of neural response similarity on location, thus on the spatial distance between the neurons, for the representation of vocalizations in the central inferior colliculus. The hypothesis is tested that despite the locally differing various input, the spectrotemporal gradients induce a gradually changing neural representation of these natural complex sounds. To this end, neural responses are compared by cross-correlation for differently and similarly frequency-tuned neurons, with respect to the spatial distance between the neurons. Simultaneous and non-simultaneous recordings were compared to obtain indications whether similarity of responses is mainly induced by the stimulus or by neuronal interactions that can induce neural correlations; the latter might suggest cooperative processing of the multi-units clusters. Whether neural correlations for vocalizations exist in the mammalian inferior colliculus has not been investigated before. Neural correlations can be beneficial, detrimental or not to affect encoding of sensory stimuli, and might depend on the specific neuronal structure \citep{AverbeckLathamReviewCorr2006}.\\
We analyzed simultaneous recordings from 32 sites in the ICC of guinea pigs in response to monaurally presented conspecific vocalizations. The set of eleven behaviorally relevant sounds \citep{BerrymanGuineaPigBehav1976} displays a wide spectrum of acoustic properties, such as amplitude and frequency modulations, harmonics and temporal correlations.\newpage It was suggested that neurons are adapted to process natural sounds \citep{NatOptRieke1995}, therefore, these might trigger responses which are not elicited by artificial or simple acoustic stimuli. Response similarity between multi-unit clusters is obtained by pairwise cross-correlation analysis of the spiking activity i.e.\ the processed output of the ICC neurons. The correlation of spiking responses is additionally compared to the correlation of long range activity, the local field potential (LFP). Multi-unit cluster activity is the combined activity mainly from neighboring single neurons that span one order of magnitude. This integrated activity could allow one to investigate local population processing in the ICC. It has also been shown that multi-unit clusters respond stronger to natural sound than single neurons \citep{TheunissenSelec2003} and that the natural stimuli can be more accurately discriminated based on these responses than based on single neuron responses \citep{EngineerDiscCortex2008}.\\
Whereas in a previous study in the mammalian ICC correlation of single neuron spike trains was investigated in dependence of the neuron's spectrotemporal properties with the finding that the best frequency is the most correlated parameter \citep{ChenNeighborSame2012} and a microcircuitry was suggested, in the present study similarity is investigated at the level of multi-unit clusters which likely display a different correlation structure. Here, natural communication sounds are used instead of artificial sounds and a wider spatial range of up to 1600 \(\mu\)m is probed. Dependencies on spatial distance have not been found for the grass frog midbrain \citep{Eggermont1987}, but have been shown  in the primary auditory cortex \citep{Eggermont2006}. Since the grass frog midbrain only displays a weak tonotopic gradient, these findings do not transfer to the mammalian inferior colliculus with a clear tonotopic organization and substantial differences in neural structure.\\
In summary, we find that neural correlations exist in the mammalian inferior colliculus and that the neural and response correlations for spiking and long range activity gradually decrease with spatial distance for similarly and differently frequency-tuned neurons. This suggests the gradual neural representation of vocalizations is shaped by interactions between the neurons and their spectral and temporal preferences.
\section{Materials \& Methods}
\subsection{Electrophysiology}
\begin{figure}[t,h!,b,p]\centering\includegraphics[width=0.85\linewidth]{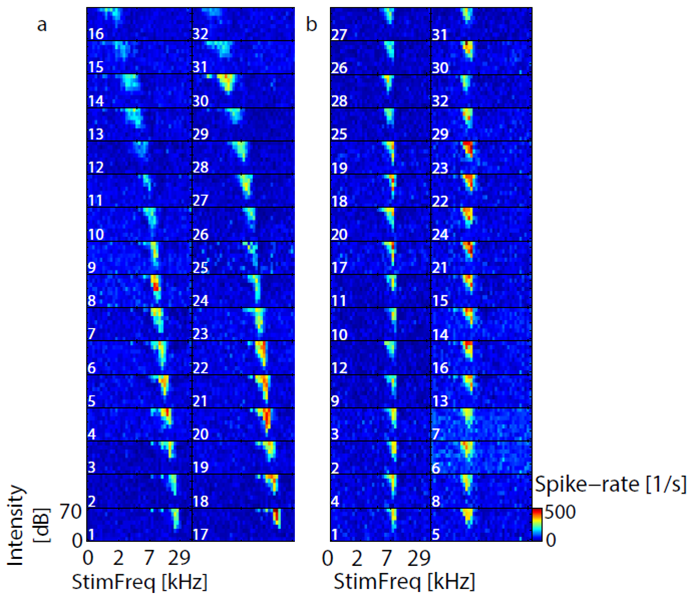}
\caption[Frequency tuning along the tonotopic and a few isofrequency axis.]{\textbf{Frequency tuning along the tonotopic gradient and within a few isofrequency lamina.} a) Frequency response maps (FRMs) recorded from 32 sites along the tonotopic gradient with a linear double-shank electrode. The characteristic frequency (CF) covers a range from 0.5-29~kHz. The CF increases gradually from sites higher up (dorsolateral) to lower ones (ventromedial), also the shape of the FRM changes from a broader symmetric shape to an elongated and skewed shape for higher frequencies. The topmost sites do not show strong responses and might be lying outside the central IC. b) FRMs of multi-units recorded with a double-tetrode electrode from two isofrequency laminae with CFs of about 7~kHz (left column) and about 4~kHz (right). The CF does not change visibly within one isofrequency lamina, but the frequency tuning varies in sensitivity, e.g.~spike rates from site 23 and site 5 vary by over 150~Hz.} \label{fig:FRM}\end{figure}

Neural activity was collected from the central nucleus of the inferior colliculus (ICC) of 11 adult male and female Dunkin Hartley guinea pigs. Recordings were acquired from 11 guinea pigs in 3 to 4 electrode insertion positions (taken altogether 36 positions), with activity recorded simultaneously from 32 recording sites. 
The electrophysiological recordings and experimental setup are described in detail elsewhere \citep{LimEnv2013,Lyzwa2016}.\\
For the recording, either a \mbox{linear double-shank} array (shank distance was 500 \(\mu\)m with 16 contacts linearly spaced at 100 \(\mu\)m, on each shank) or a \mbox{4-double-tetrode} array (shank distance of 500 \(\mu\)m, contact distance within a tetrode of 25-82 \(\mu\)m) were used to measure activity simultaneously from 32 sites (impedances were 0.5-1 M\text{$\Omega$} at 1 kHz; NeuroNexus Technologies, Ann Arbor, MI). Whereas the linear-double shank array mainly records responses along the best frequency gradient, and covers a broad range of best frequencies, the 4-double-tetrodes record from a few neighboring isofrequency laminae, and several multi-unit clusters have similar frequency tuning (Fig.\ref{fig:FRM}). 
They have similar best frequencies, but might have different preferences for amplitude modulations (AM) depending on their spatial distance within the ICC \citep{LangnerPeriod1988b, Rees2011}. 
The electrode array was introduced under an angle of 45$^\circ$ dorsolaterally along the tonotopic gradient of the ICC, into 3 to 4 different insertion positions for each animal. The animals were anesthetized with Ketamine and stereotactically fixed with ear tubes through which the sound was presented directly to the left eardrum. While acoustically presenting vocalization stimuli to the left ear, neural activity was recorded from the contralateral ICC at a sampling rate of 24.414~kHz using a TDT Tucker Davis System. For each vocalization 20 trials were recorded with intensities of 30-70~dB~SPL in steps of 10~dB~SPL. For the analysis of this study recordings are those at 70~dB SPL stimulus intensity, as these show the strongest response.\\
Frequency response maps (FRM) were obtained from spiking responses to pure tone stimuli. A total of 40~stimulus frequencies, ranging between 0.5 to 45~kHz, with a ramp rise and fall time of 5~ms each and a duration of 50~ms were presented. From the FRMs, the best and characteristic frequencies were obtained. These tone-evoked preferred frequencies, the frequency eliciting the highest spike-rate at the lowest spike-eliciting intensity (the characteristic frequency, CF) and at a given presented stimulus intensity (the best frequency, BF) ranged from 0.5 to 45~kHz. The frequency response maps for a linear double-shank recording along the tonotopic gradient and for a 4-double-tetrode recording are given in Fig.~\ref{fig:FRM}.
\subsection{Vocalization Stimuli}
\begin{figure}[t,b,p]\centering\includegraphics[width=0.98\linewidth]{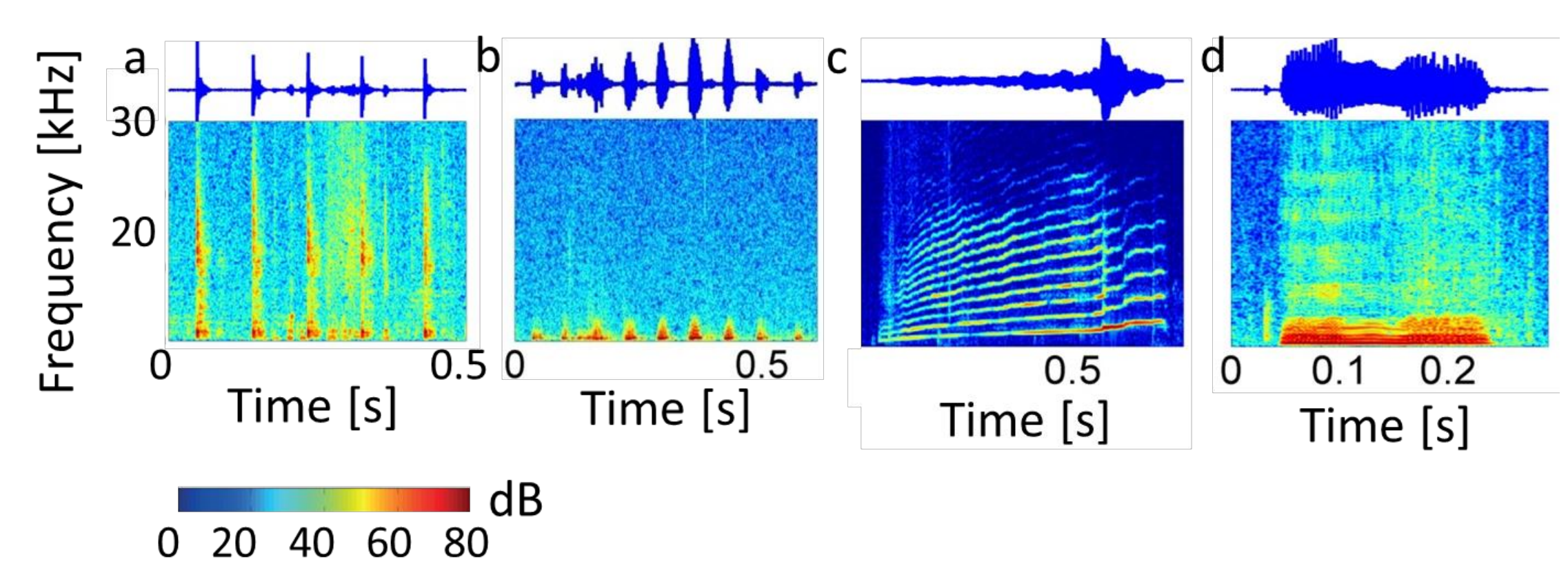}\caption[Example vocalizations.]{\textbf{Vocalizations} Spectrograms and waveforms of four representative examples of the entire set of eleven guinea pig vocalizations. The `tooth chatter' (a) and `drr' (b) have periodicities in the waveform, with (a) containing frequencies up to 30 kHz, and (b) containing only low frequencies below 3 kHz. The `scream' (c) and `squeal' (d) have complex waveforms, cover a broad spectral range and show harmonics.}\label{fig:vocs4}\end{figure}

The 11 vocalization stimuli used in this study are a representative set of guinea pig communication calls and give information about the animal's behavioral state \citep{BerrymanGuineaPigBehav1976}. Figure \ref{fig:vocs4} shows the spectrograms and waveforms for four examples of this set: the `tooth chatter', `drr', `scream' and `squeal'. These natural complex sounds display a variety of frequency modulations, frequency ranges and envelope types. Some vocalizations such as the `drr' (b) contain mainly frequencies below 3 kHz and have periodicities in the waveform. The `tooth chatter' also has a periodic waveform but a frequency content of up to 30 kHz. Others such as the `scream' (c) and `squeal' (d) have complex waveforms, cover a broad spectral range and display harmonics. Vocalizations with a complex waveform and almost all energy at low frequencies are also present. \newpage The vocalizations were played 20~ms after recording onset and vary in duration between 300~ms and 1,300~ms. They were recorded from male and female Dunkin Hartley guinea pigs at a sampling rate of 97.656~kHz. Details on the stimuli recording including sound calibration and the frequency content for the full vocalization set can be found in \citep{LimEnv2013, Lyzwa2016}.

\subsection{Preprocessing of Neurophysiological Data}
In order to investigate responses to vocalizations in the ICC, the spiking activity is employed for the analysis, because this is the processed output of the neurons \citep{GauteLFP2012}. Additionally, the analysis is extended to the local field potential (LFP), which is long-range activity and contains i.a.\ the synaptic input to the neurons \citep{GauteLFP2012}. The local field potentials were obtained by Butterworth filtering the voltage traces in a range between 0.5-500~Hz \citep{GauteLFP2012}.\\
To obtain spiking multi-unit activity the recorded voltage traces are Butterworth-filtered for 300-3,000~Hz and thresholded at \(z =3\) standard deviations exceeding the ongoing activity (\(\Theta=\mu +z\sigma\)), with the mean \(\mu\) and the standard deviation \(\sigma\) of the ongoing activity. This spontaneous or ongoing activity was acquired from the first 20~ms of each recording, during which no stimulus was presented, in order to account for adaptation effects over time and different spontaneous rates of the neurons. Due to the low impedance of \(0.5-1 M\text{$\Omega$} \), detected spikes likely originate from different single neurons, and therefore no refraction time between spikes was assumed. The recorded activity is multi-unit cluster activity, which is the compound spiking response mainly of several neighboring single neurons recorded from one site. We used the offline spike-sorting program \textit{WaveClus} \citep{QuirogaWaveclus2004} to sort and separate spikes according to the spike waveform on a subset of the passband-filtered multi-unit cluster responses, with details in \citep{Lyzwa2016}. Separation into single units was not possible because within sorted clusters a significant fraction of the spikes displayed inter-spike-intervals of less than 3 ms; and in some cases spikes could not be sorted into clusters. The responses investigated here are from neural groups comprising at least 3-5 single neurons and smaller contributions from neurons that are farther away from the recording electrode and are not distinguishable. Note that it is possible that different sub-groups of multi-unit clusters respond to different vocalizations. Multi-unit spike trains were binned at 1~ms and convolved with an exponential filter function, \(\mathit{f(t)}{=}\mathit{t}\cdot \mathrm{exp}{\mathrm{(\alpha}\cdot\mathit{ t)}}\), with time \(\mathit{t}\), to mimic the time course of excitatory postsynaptic potentials (EPSP) \citep{RossumMetric2001}, as used by \citep{MachensSUdisc2003}. The full width at half maximum \(\mathrm{\alpha}\), of the EPSP-like function was chosen to be 3~ms \citep{Lyzwa2016}. Averaging the binned spike trains across all \(n_{\mathrm{trial}}\,=\,\)20 trials yielded the post-stimulus time histogram (PSTH).
\subsection{Cross-Correlation Analysis}
In order to test similarity of responses from different multi-unit clusters to the same vocalization, we employed cross-correlation which yields a compact description for the large set of neurons analyzed in this work. Similarity of responses to each vocalization was tested for pairs of multi-unit clusters from one recording which allows including the spatial distance between the neurons into the analysis. 
A correlation-based similarity measure of spike trains \citep{Schreiber2003} has been employed earlier for neural discrimination of single neurons and groups of neurons \citep{Wang2007}. It has been shown that for the comprehension of speech \citep{ShannonTemp1995} which is spectrotemporally varying complex sound as are vocalizations, temporal information is crucial. As an index of the responses' temporal information, here, we computed the degree of correlation between the spike trains from different multi-unit clusters. Pairwise, EPSP-spike trains or LFP from two multi-unit clusters \(\mathit{ x(t), y(t)}\) of length \( \mathit{n} \), were cross-correlated and the highest value within a maximum possible delay of \(\tau\) between the responses was selected: \begin{equation} \mathit{\mathrm{Corr(\tau)}} = \mathrm{max} \left(\mathit{ \frac{ \sum_{t = 1}^{n -\tau} (x(t+\tau)-\langle x\rangle) \cdot (y(t)-\langle y\rangle)}{\sqrt{\sum_{t = 1}^{n} (x(t)-\langle x \rangle)^2 \sum_{t = 1}^{n}(y(t)-\langle y\rangle)^2}}} \right) \end{equation} with a lag of \( \mathrm{\tau} =\left[ -\mathrm{10\  ms},\mathrm{10\  ms} \right] \). 
This delay is within the range of maximum response latencies in the ICC \citep{LangnerLatency1987}. The correlation values were computed with a lag, because response latencies do vary across multi-unit clusters with different spectral preferences \citep{LangnerLatency1987}. The correlation value of \(\mathit{n_{\mathrm{trial}}} = 20\) trials for this multi-unit pair was then averaged. 
\subsection{Neural Correlations}
Simultaneously recorded spiking responses from interacting neurons can show correlated trial-to trial variability \citep{AverbeckLathamReviewCorr2006}. If no stimulus is present the correlated activity is termed neural or noise correlation (\(N\)). For stimulus-driven simultaneous responses, the measured response correlation contains \(\mathrm{Corr}\), in addition to the neural correlation, a contribution which is due to the neurons responding to and following the same stimulus (stimulus correlations, \(D\)). These stimulus correlations are present for simultaneous and non-simultaneous recordings, as long as the same stimulus is used.
In order to separate the stimulus from the response correlations, simultaneously recorded trials of the multi-unit clusters were randomly shuffled \citep{AbelesShiftPred1982} before correlating them, and thus only stimulus correlations remain. 
To investigate whether neural correlations exist and to obtain them, the averaged stimulus correlation is subtracted from the response correlation for each trial and the average is taken \(N =\sum_{i}^{n=20} \mathrm{Corr(i)}-S(i)\).\\ 
This approach \citep{AbelesShiftPred1982} has been developed for single neurons, and attempts to infer functional connectivity from the computed neural correlations. Here, we use multi-unit clusters, and we do not attempt to make inferences about functional neural connections but to test whether neural correlations exist. If correlated activity is present then there is a significant difference between response correlations and the stimulus correlations which is likely due to neural interactions. 
Significance was assessed using the Student's \(\mathit{t}\)-test for normal distributions, and the Wilcoxon-Mann-Whitney test for comparison of non-normal distributions.\\ 
In order to visualize the effect of noise correlations on the encoded stimulus information, scatter plots are often used \citep{AverbeckLathamReviewCorr2006}. They display the distribution of averaged spike rates of two neurons or neural groups to different stimuli. The more the response distributions to the different stimuli overlap, the less information is carried by them. Comparing the scatter plot of simultaneous trials to the one with shuffled trials yields information whether the neural correlation affect encoding. If, for example, separability of responses to different stimuli increases when removing neural correlations, these are detrimental for encoding.
\subsection{Frequency Tuning and Spatial Distance of Neuronal Pairs}
Correlation is investigated for neuronal pairs of two multi-unit clusters. The two multi-unit clusters are characterized by their spectral and relative spatial distance. 
The spectral distance is the absolute difference between the characteristic frequencies of the two multi-unit clusters. Pairs differing by more than 1\,/3~octave in their characteristic are most probably from different isofrequency lamina along the tonotopic gradient \citep{LangnerPeriod1988b}. These pairs were assigned to the group of differently frequency-tuned neurons. Those pairs that have the same preferred frequency within an interval of 1\,/3~octave are likely from the same isofrequency lamina \citep{LangnerPeriod1988b} and were assigned to the group of similarly frequency-tuned neurons.\\
The spatial distances between the multi-unit clusters of all pairs were mapped according to the channels on the electrode. Distances between all 32 channels were either obtained directly from the NeuroNexus manual (NeuroNexus Catalog, Ann
Arbor, MI 48108, 2014) or calculated using the Pythagorean theorem, yielding a \(32\,\times\,32\)~matrix, respectively for the linear double-shank and the 4-double-tetrode electrode array. Minimum and maximum distances for the double-shank and double-tetrode array were respectively \(\mathit{D_{\mathrm{min}}\,=}\)100~\(\mu\)m, \(\mathit{D_{\mathrm{max}}\,=}\)1581~\(\mu\)m and \(\mathit{T_{\mathrm{min}}\,=}\)25~\(\mu\)m, \(\mathit{T_{\mathrm{max}}\,=}\)1372~\(\mu\)m.\\ 
The correlation analysis was performed for pairs from the same recording, yielding ideally \mbox{(\(32\times31)/2\,=\,\)496 pairs, which do not include autocorrelations or correlations counted twice.} However, only pairs with a minimum distance of 200~\(\mu\)m were considered for the analysis of similarly tuned neurons (respectively 100~\(\mu\)m for differently tuned neurons) \citep{RecordingRangeBuzaki2004, MalmiercaRess1995}, in order to assure that no multi-unit clusters are taken from adjacent recording sites, which possibly share the same neurons and therefore yield excessive high correlation values. 
The distribution of spatial distances was not uniform and also varied for double-shank and tetrode recordings. Therefore, the number of multi-unit clusters, for which correlation values were averaged for one spatial distance, varied across recordings. Differences in response similarity based on different amplitude modulation preferences for multi-unit clusters with the same frequency tuning might be averaged out when taking the mean across multi-unit pairs. Note that the use of multi-unit clusters for the study could limit the ability to assess the degree to which correlated firing may encode temporal features in the vocalizations.\\
Correlation is computed between multi-unit responses to the same vocalization, from one recording set which consists of responses from 32 multi-unit clusters. The analysis is repeated for all 36 recordings and analyzed for each recording. It was verified that the observed trend is consistent across recording sets. In the following, results for an individual example recording set are shown and the average across all recording sets (1,152 multi-unit clusters) is displayed. When averaging values across multi-unit pairs, the displayed error bars were chosen to represent one standard deviation to indicate correlation variability across multi-unit pairs. The error of the correlation values were computed via error propagation and are minor. 
\section{Results}
\noindent
We analyzed response and neural correlation from 1,152 multi-unit clusters across a wide frequency range of the central inferior colliculus of 11 guinea pigs for a spectrotemporally rich set of 11 species-specific vocalizations. Using cross-correlation, for spiking and LFP activity, we tested variation of similarity for individual vocalizations across the ICC and investigated whether correlation values depend on the spatial distance between multi-unit cluster. We compare response correlations and the contributions respectively due to the stimulus and due to neural interactions. At first, we display time-averaged neural responses to vocalizations, and then show correlations for similarly and differently tuned neurons.\newpage
\begin{figure}[t,h,b,p]\centering\includegraphics[width=0.82\linewidth]{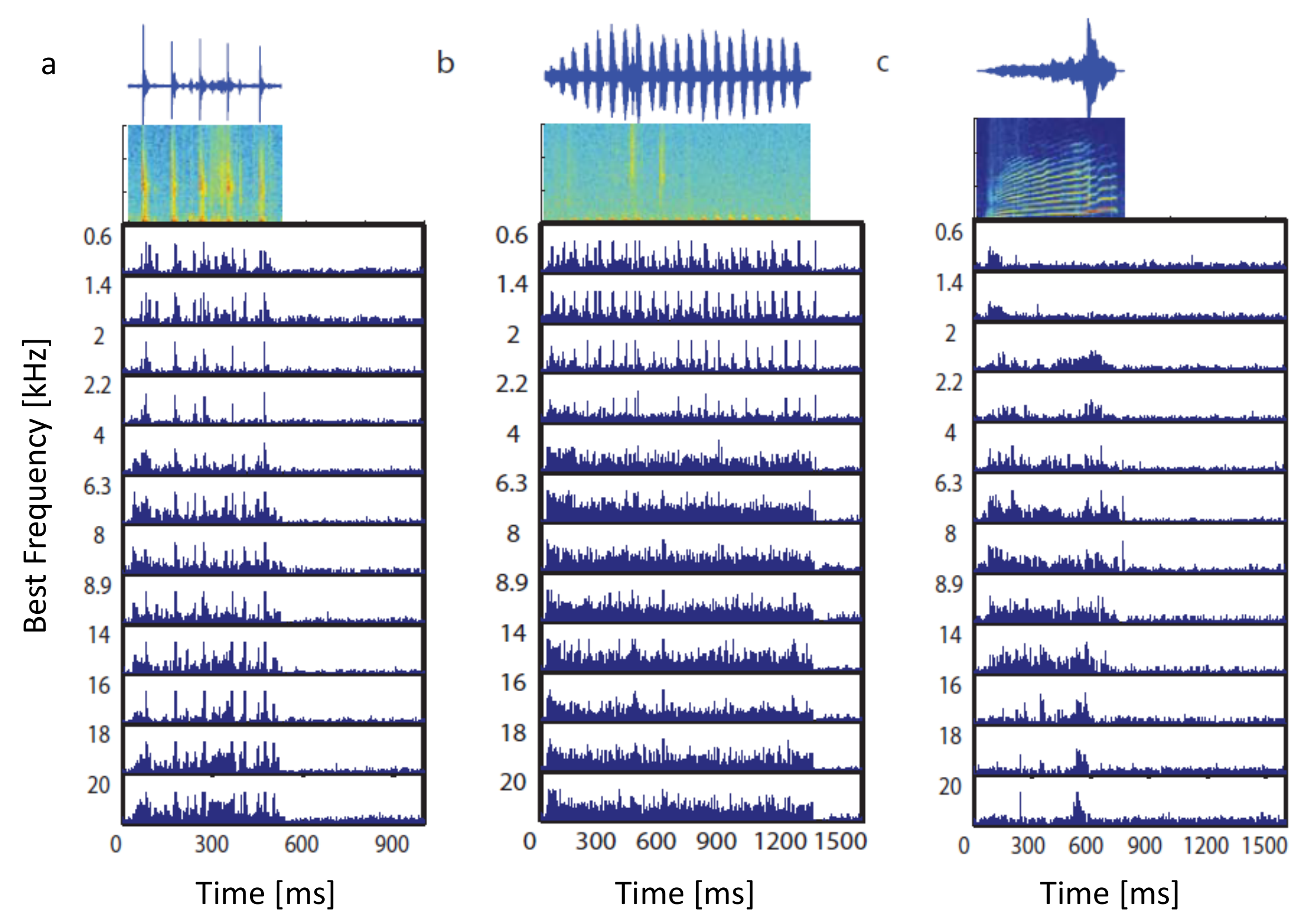}
\caption[Post-stimulus time histograms of vocalizations.]{\textbf{Vocalization Post-stimulus time histograms.} PSTHs in response to the vocalizations a)~`tooth chatter',~b)~`purr', and c)~`long scream'; for multi-unit clusters from of a linear double-shank recording, spanning a best frequency range of 0.6-20~kHz. On top, the waveforms and spectrograms are displayed.}\label{fig:psth}\end{figure}
\subsection{Neural Responses}
The post-stimulus time histogram (PSTH) represents the trial-averaged (\(\mathit{n_{\mathrm{trial}}}=\)20) temporal neural response. Responses to vocalizations vary for differently frequency-tuned multi-unit clusters and follow the spectrally matching components in the stimulus. 
Figure~\ref{fig:psth} displays the PSTHs of multi-unit clusters along the best frequency (BF) gradient in response to three vocalizations. Responses to the `tooth chatter' phase-lock to the stimulus envelope for multi-unit clusters throughout the whole best frequency range (Fig.~\ref{fig:psth}a), because the stimulus has spectral energy in this range (Fig.~\ref{fig:vocs4}a). However, the response in general becomes broader for high frequency neurons. Responses to the `purr', on the other hand, phase-lock accurately only for low frequency neurons, but then become broad and unspecific (Fig.~\ref{fig:psth}b). Spectral energy is present for frequencies up to 3~kHz in the `purr' vocalization. 
At the start and the end of the stimulus presentation, onset and rebound responses are more pronounced for middle and high best frequency neurons. The responses' dependence on the match of the best frequency and the spectral content of the stimulus \citep{Suta2003} is clearly illustrated by responses to the `scream' (Fig.~\ref{fig:psth}c). In the beginning of the stimulus, only low frequencies are present, and only low-BF multi-unit clusters respond. Subsequently, the stimulus contains frequencies up to 25~kHz and middle-BF multi-unit clusters respond. High-BF multi-unit clusters respond to a high frequency peak at 600~ms. In some cases, these multi-unit responses can be approximated by the bandpass filtered waveform of the vocalization, filtered around the best frequency of the multi-unit cluster \citep{Lyzwa2015}, hence it would be likely that clusters with the same best frequency also have higher correlations of their responses. Similarity of responses to the same stimulus by different multi-unit clusters can be directly obtained by comparing the PSTHs. In order to quantify similarities across the large data set of multi-unit clusters used in this work, a more compact measure is employed. To this end, responses are cross-correlated and the correlation value indicates the degree of response similarity.
\subsection{Dependence on Similarity of Frequency Tuning}
\begin{figure}[t,b,p]\centering	
   \includegraphics[width=0.99\linewidth]{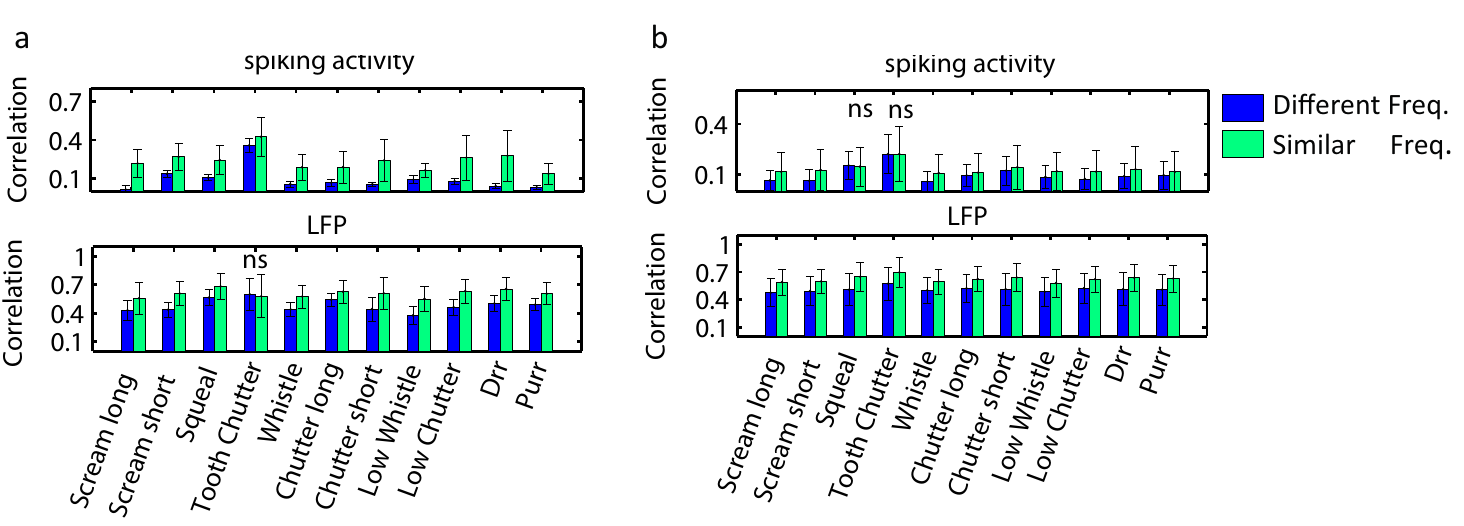}
\caption[Correlations within and across isofrequency laminae.]{\textbf{Correlations for similarly and differently frequency-tuned neurons.} Averaged correlation values for multi-unit pairs a)~from one recording (n\,\(=\)\,155), and b)~averaged across all recordings. For comparison, the number of pairs was kept constant for the two cases, n\,\(=\)\,4223. Top: Spiking activity, bottom: local field potentials. Responses for similarly frequency-tuned neurons are mostly significantly more correlated than for differently frequency-tuned neurons (two-sided Wilcoxon-Mann-Whitney test \(\mathit{p}\)\,\(=\)\,0.05).}
\label{fig:TonoIso}
\end{figure}

Responses correlations to vocalizations are compared for multi-unit cluster with similar and different frequency tuning.\newpage  In general, neurons within the same isofrequency lamina have similar preferred frequency (within 1/3 octave \citep{LangnerPeriod1988b}) but possibly different preferences for amplitude modulations \citep{LangnerPeriod1988b, Langner2002, Rees2011}. Neurons with different frequency tuning properties are in different laminae and along the tonotopic gradient. 
As an example for multi-unit pairs from only one recording, Figure~\ref{fig:TonoIso}a displays the correlation values for each vocalization, for spiking activity and local field potentials. Correlation values show large variability within one recording set, as depicted by the error bars which correspond to one standard deviation. However, correlation values are significantly higher for similarly frequency-tuned pairs than for neurons with different best frequencies as assessed by the two-sided Wilcoxon-Mann-Whitney test, \(\mathit{p}\)\,\(<\)\,0.05, for all vocalizations (except for the `tooth chatter' in LFPs). These correlation values vary across the different recording sets, because of the different distributions of frequency tuning similarity and spatial distances between the neuron pairs.
For the average correlation values across all recordings (n\(\,=\,\)4223 multi-unit pairs), differences in the correlation are significant for all vocalizations for LFP activity, Fig.~\ref{fig:TonoIso}b. Differences in correlation for similar and different frequency tuning are smaller for spiking activity than for LFP and not significant for the `squeal' and `tooth chatter', Fig.~\ref{fig:TonoIso}b. The `tooth chatter' shows strong phase-locking across the tonotopic gradient, the `squeal' displays partly spectral energy across frequency bands and therefore also a neural response across frequency bands. The variability of correlation across multi-unit pairs, which is large, is depicted by the error bars. The errors computed via error propagation are very small. Significance is evaluated by the two-sided Wilcoxon-Mann-Whitney test, \(\mathit{p}\)\,\(<\)\,0.05.\\
Correlation values are significantly higher for LFP than for spiking activity for each vocalization, for similarly and differently frequency-tuned neurons (\(\mathit{p}\)\,\(<\)\,0.05). The LFP responses are long-range activity, and spread throughout wider spatial and frequency regions than spiking responses which are confined to the range of one multi-unit cluster. Across vocalizations, response similarity varies significantly, except for (2,4;9,10), respectively for spiking (LFP) responses for similar and different neural frequency preference. The responses to the `tooth chatter' are significantly larger than to all other vocalizations, which is due to the stimulus' spectral content across a wide frequency range and the responses' phase-locking throughout this range (see Fig.~\ref{fig:psth}a).\\
Higher correlation within similar frequency regions than across the frequency gradient might point to vocalizations being processed by isofrequency laminae as functional units \citep{LangnerLaminar1997}, however, even though the differences are significant, they are minor and variability exists across multi-unit clusters. In the following, correlation dependence on spatial distance will be investigated separately for these two groups. 
\begin{figure}[t,h!,p]\centering\includegraphics[width=0.75\linewidth]{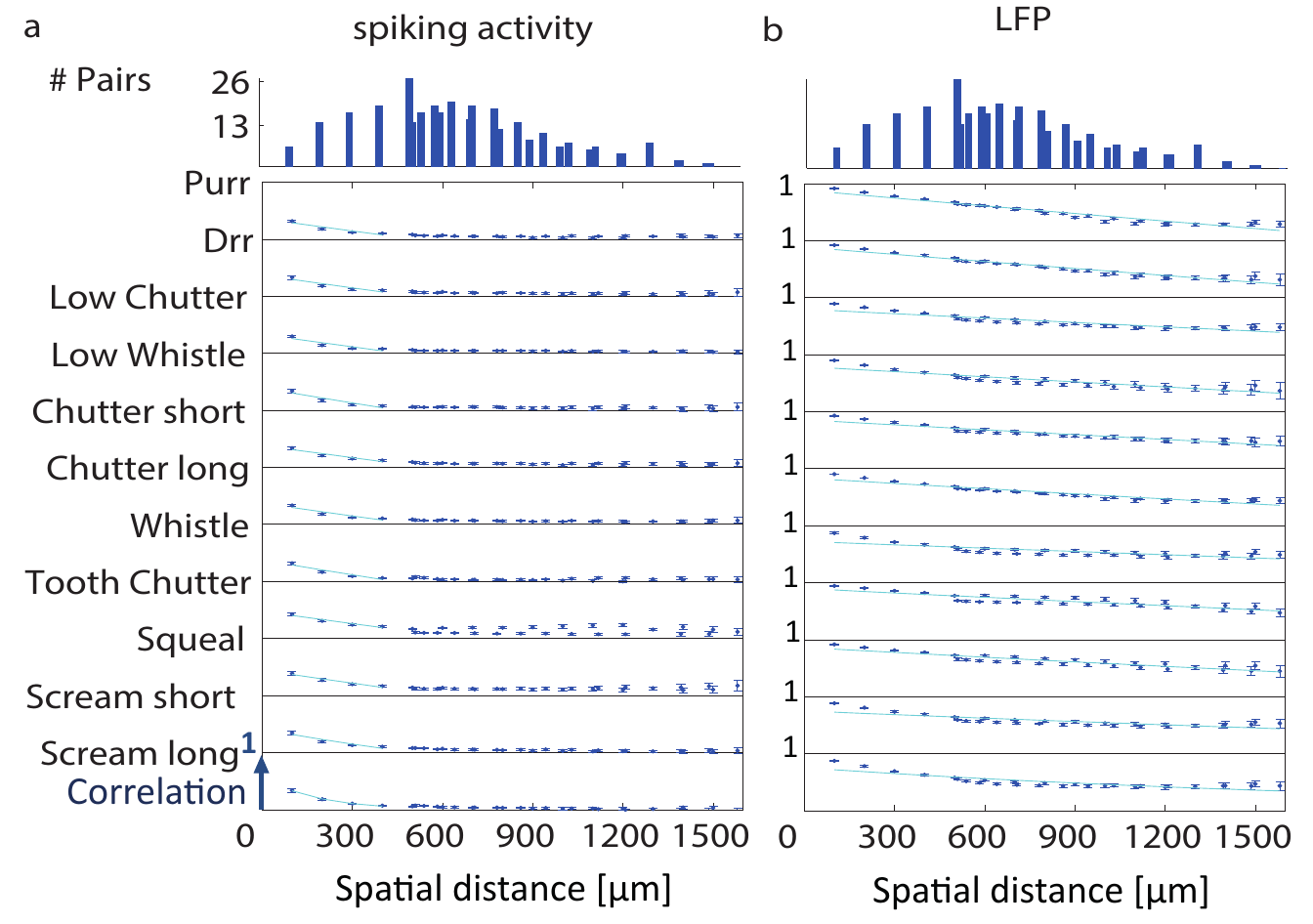} \caption[Correlation dependence on spatial distance for differently frequency-tuned multi-unit cluster.]{\textbf{Correlation dependence on spatial distance for differently frequency-tuned neurons.} Averaged correlations values for each distance from one recording set are displayed for all vocalizations, for a)~spiking activity, and b)~Local field potentials. Correlations decrease almost linearly with spatial distance, with a much bigger decrease for spiking activity than for LFP. Exponential regression of this decrease (\textcolor{cyan}{\textbf{--}}), with overlaps of 61-100 \(\%\), see Tab.~\ref{tab:fit}. On top, the number of multi-unit pairs for each spatial distance over which the average correlation was computed is displayed. The y-axis for each vocalization shows a correlation range of 0-1. }\label{fig:LinTono1}\end{figure}
\subsection{Dependence on Spatial Distance}
\begin{figure}[t,h!,b,p]\centering\includegraphics[width=0.60\linewidth]{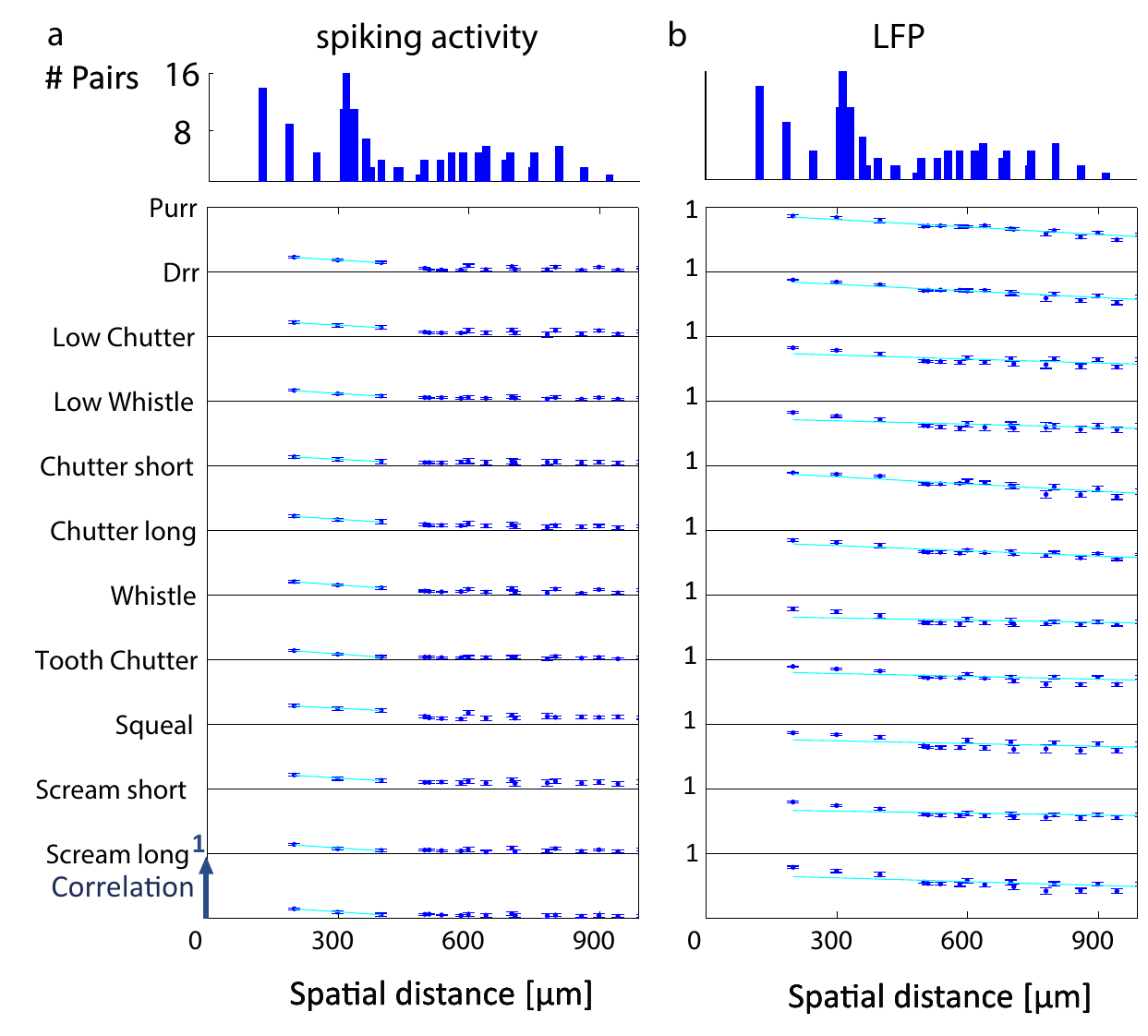}
\caption[Correlation dependence on spatial distance for similarly frequency-tuned multi-unit cluster.]{\textbf{Correlation dependence on spatial distance for similarly frequency-tuned neurons.} Averaged correlations values for each distance from one recording set are displayed for all vocalizations, for a)~spiking activity, and b)~LFPs. Exponential regression of this decrease (\textcolor{cyan}{\textbf{--}}), with overlaps of 41-100 \(\%\) see Tab.~\ref{tab:fit}. On top, the number of multi-unit pairs for each spatial distance over which the average correlation was computed is displayed. Distances 200 \(\mu\)m-1000 \(\mu\)m for similarly tuned neurons are analyzed, see 2.2. } \label{fig:LinIso1}\end{figure}

\begin{figure}[b,h!,t,p]\centering\includegraphics[width=0.80\linewidth]{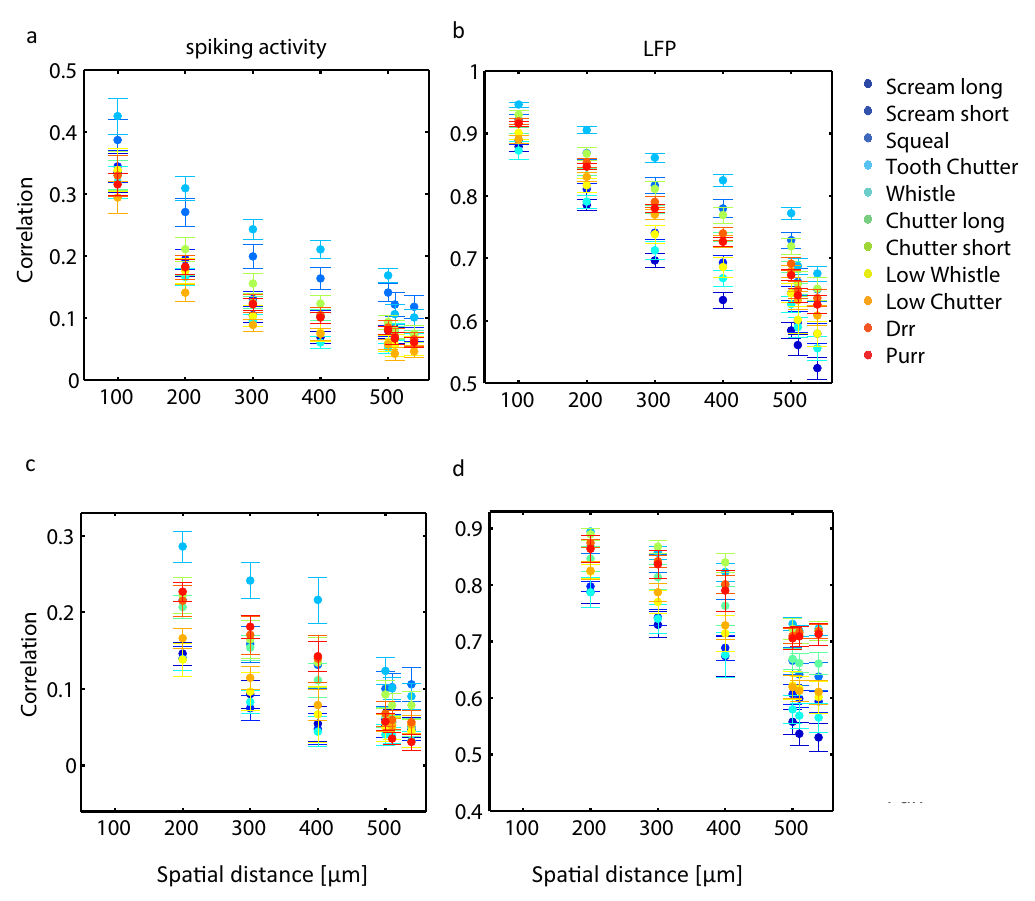}
\caption[Comparison of spatial dependence for all vocalizations.]{\textbf{Comparison of correlations for all vocalizations.} Correlations are shown for differently (a,b) and similarly (c,d) frequency-tuned neurons for spiking activity (a,c) and LFPs (b,d), displayed in Fig.~\ref{fig:LinTono1},~\ref{fig:LinIso1}. Differences in total value of response correlations exist across vocalizations. For each vocalization all correlations display an almost linear decrease with spatial distance between neurons.}\label{fig:LinVoc}\end{figure}

In order to display the relation between correlation values and the spatial distance between neurons, the values were averaged for multi-unit clusters for each spatial distance of one recording set. The relations are shown for spiking activity and LFPs, an example for one recording set of pairs with different (Fig.~\ref{fig:LinTono1}) and similar frequency tuning (Fig.~\ref{fig:LinIso1}) is given. Correlations decrease with spatial distance and are almost zero for distances of 400~\(\mu\)m for the spiking responses. LFP correlations are overall higher than the ones for spiking activity and display a less rapid decrease with distance, correlation values of about 0.5 still exist for the maximum measured distance of 1600~\(\mu\)m (Fig.~\ref{fig:LinTono1}). LFP is long range activity and correlations are present over long distances. For multi-unit clusters from one recording set, that have similar frequency tuning this decrease is also observed (Fig.~\ref{fig:LinIso1}). These findings are consistent across all 36 analyzed recording sets. Neural and stimulus correlations (3.4) follow the same decrease as the response correlations.\\
The decrease can be approximated with an exponential function \(\mathit{f(x)=a\cdot e^{-bx},}\) with \(x\) the spatial distance. In Tab. \ref{tab:fit} values for \(\mathit{a,~b}\) and the match of data and fit \(\mathit{\chi}\) are given.\\
\begin{table}[h!]\begin{center}
\begin{tabular}{|cccccc|}\hline 
\text{Activity} & \text{Frequency Tuning} &\text{a}& \text{b}\( \left[ 1/\mu \mathrm{m}\right]\) & \(\chi \left[ \% \right] \)  & \(\  \left\langle \chi \right\rangle \left[ \% \right] \)  \\
\hline
\text{spike}  & \text{different}       &     0.01-0.59     &    0.57 -0.0069    & 84-100     &  91   \\
\text{LFP}     & \text{different}       &     0.76-1          &    0.012-0.003     & 61-94       &  80   \\
\text{spike}  & \text{similar}          &     0.01-0.39     &    0.48-0.0041     & 95-100     &  98   \\
\text{LFP}     & \text{similar}          &     0.73-1          &    0.042-0.002     & 42-93       &  67   \\
\hline 
\end{tabular}
\caption{\label{tab:fit}\textbf{Parameter for exponential regression.} \(\mathit{f(x)=a\cdot e^{-bx}}\), with spatial distance x, for Fig.~\ref{fig:LinTono1}a,b, Fig.~\ref{fig:LinIso1}a,b.}\end{center}\end{table}\\
Dependencies on the spatial distance differ across vocalizations. Figure~\ref{fig:LinVoc} shows the correlations displayed in Fig.~\ref{fig:LinTono1},~\ref{fig:LinIso1} in a smaller window for distances up to 650~\(\mu\)m for all eleven vocalizations for spiking and LFP activity. The `tooth chatter' shows overall highest response correlations. A clear correlation decrease is observed from 200-500~\(\mu\)m for differently (Figure~\ref{fig:LinVoc}a, c) and similarly (~\ref{fig:LinVoc}b, d) frequency-tuned neurons for all vocalizations.\newpage
 \newpage
Similarity of responses decreases almost linearly with spatial distance, and is almost zero for distances above 400 \(\mu\)m for spiking activity, whereas LFP shows stronger and longer range correlations even for the maximum measured distance.The parameter for the decrease are within the same range for similarly and differently frequency-tuned neurons, but the decrease is much smaller for LFP than for spiking activity (Tab.~\ref{tab:fit}). This gradual decrease within similar and different frequency regions suggests that the neural response is strongly influenced by the gradually organization of spectral and temporal preferences in the ICC.
\subsection{Neural Correlations}
\begin{figure}[h!,t,b,p]\centering\includegraphics[width=0.90\linewidth]{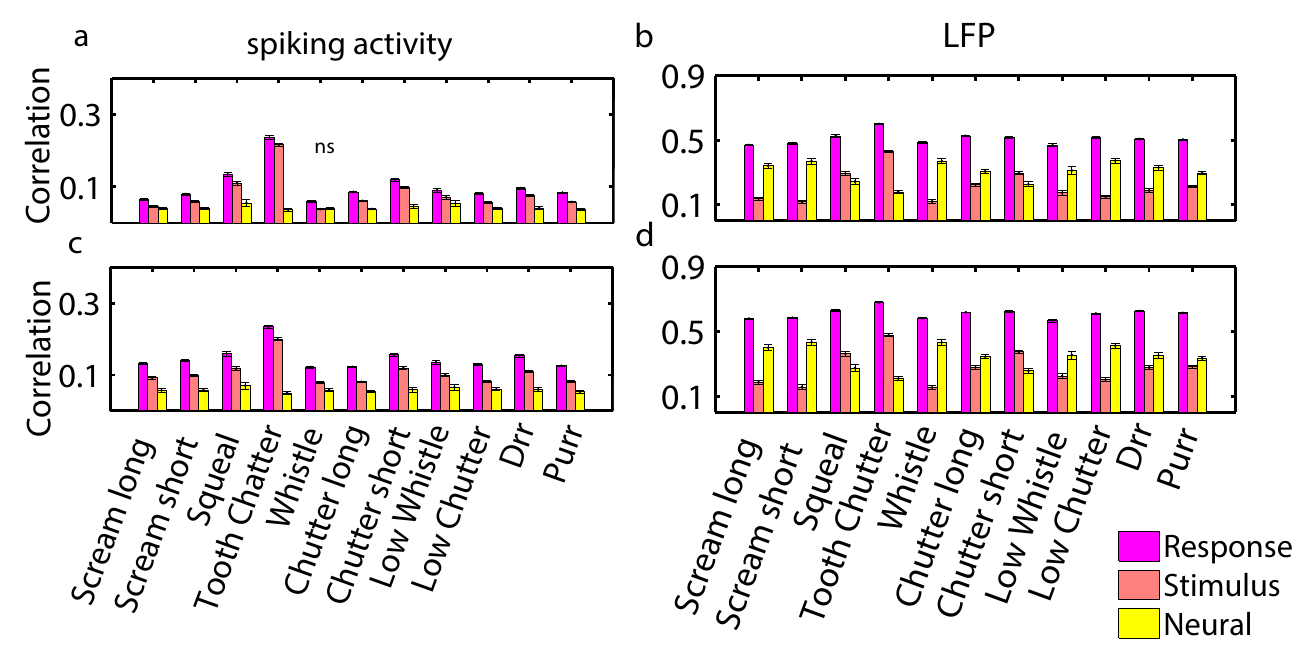}\caption[Averaged response, stimulus-driven and neural correlations.]{Response, stimulus-driven and neural correlations, averaged across all recordings. Correlations for spiking activity, for (a) differently (n\(=\)10,775 pairs); and (c) similarly frequency-tuned neurons (n\(=\)4230 pairs); and correlations for LFP activity for (b) differently, and (d) similarly frequency-tuned neurons. Non-significant differences are denoted by \textbf{ns}, all other differences are significant (\(\mathit{p}\)\,\(=\)\,0.05).}\label{fig:ShuffleCorrAll}\end{figure}

Comparison of correlations from simultaneous responses to those of non-simultaneous responses gives an indication of the amount of correlations due to neural interactions between the multi-unit clusters. It has been shown previously that the sum of the stimulus-driven and neural correlations does not necessarily yield the response correlation \citep{MelssenNeuralConnec1987}.\\
Figure~\ref{fig:ShuffleCorrAll} displays all three types: the averaged response correlations, the correlations due to the stimulus and the neural correlations, for multi-unit pairs with either similar or different frequency tuning, both, for spiking (a) and LFP (b) activity. The displayed error was obtained via error propagation. Correlation values vary across multi-unit clusters as displayed in Fig.~\ref{fig:TonoIso}.\\
Response correlations of the spiking activity are significantly larger than non-simultaneous correlations, and these are larger than the neural correlations. The difference between response, stimulus-driven and neural correlations is significant for all vocalizations for spiking and LFP activity (Fig.~\ref{fig:ShuffleCorrAll}). Only the `whistle', which has the overall lowest response correlations, does not show significant difference between the stimulus and neural correlation of spiking activity for differently frequency-tuned multi-unit pairs, Fig.~\ref{fig:ShuffleCorrAll}a, top. Differences between response and stimulus correlations are not significant for all vocalizations within each recording set, but are significant when averaged across all multi-unit cluster.

Differences between response, stimulus-driven and neural correlations are larger for LFP than for spiking activity. These are significant for correlations of all vocalizations in single recording sets and for averaged values across all multi-unit clusters. For the LFP activity, neural correlations are even significantly higher than stimulus correlations for all vocalizations but three, for both, similarly and differently frequency tuned multi-unit pairs (Fig.~\ref{fig:ShuffleCorrAll}b).\newpage The three exceptions, the `tooth chatter', `short chutter', and `squeal', show energy across the whole frequency range at short time intervals in the vocalization spectrograms, Fig.~\ref{fig:vocs4} and \citep{Lyzwa2016}. For the spiking activity, these three vocalizations also show the overall smallest amount of neural correlations relative to the response correlations.\\
To summarize, neural correlations which are most likely due to interactions between multi-unit clusters exist in the central inferior colliculus for spiking activity and for local field potentials. These correlations exist between similarly and differently frequency-tuned multi-unit activity.
They are significant but minor for spiking activity, and much larger for LFPs which is long range activity \citep{GauteLFP2012}. Thus, the responses are also shaped by the interactions between neurons. \\

\begin{figure}[h!,t,b]\centering\includegraphics[width=0.99\linewidth]{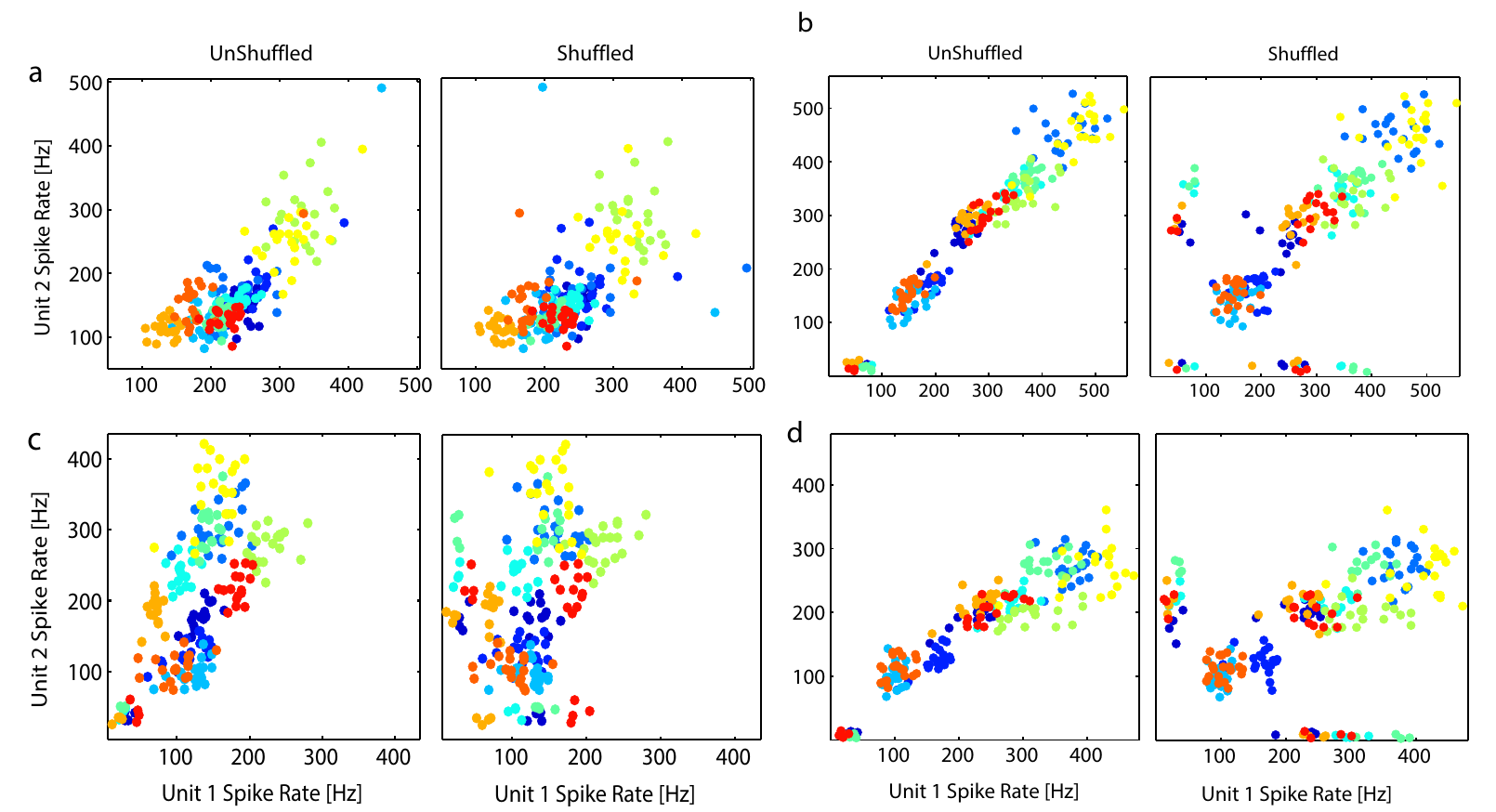}\caption[Separation of spike rates with correlated trial-variability.]{\textbf{Separation of spike rates with correlated trial-variability.} The scatter plots of the average spike rates in response to the 11 vocalizations for simultaneous (UnShuffled) and non-simultaneous (Shuffled) trials are compared for 4 multi-unit pairs, for units with different (a,~c) or similar frequency tuning (b,~d); and which are spatially relatively close (a,~b) or far away (c,~d). \mbox{a)~\(\mathrm{BF_{1,2}\,=\,5~kHz,~8~kHz;~360~\mu m}\);}\mbox{b)~\(\mathrm{BF_{1,2}\,=\,2.25~kHz,~2.83~kHz;~400~\mu}\)m;}\\
\mbox{c)~\(\mathrm{BF_{1,2}\,=\,8~kHz,~2.25~kHz;~1270~\mu m}\);}\mbox{d)~\(\mathrm{BF_{1,2}\,=\,2~kHz,~2.25~kHz;~800~\mu m}\).} The colors for the different vocalizations correspond to those in Fig. \ref{fig:LinVoc}.}\label{fig:noiseCorr}\end{figure}
Neural correlations exist in the inferior colliculus. Interactions between neurons can lead to a co-variation of their trial-to-trial variability of the spiking responses, which might also manifest in a correlated trial-variability of their spike-rates. The averaged stimulus-elicited spike rate for each trial (n\(=\)20), for multi-unit pairs are compared for simultaneous and non-simultaneous responses to investigate whether a decorrelation (shuffling) induces better separability of the spike rates to different vocalizations. Thus, we explored whether the contribution of neural correlations to separate spike rates of different vocalizations differed among multi-unit pairs with a small or large difference either in frequency tuning, or in spatial distance. Figure~\ref{fig:noiseCorr} displays examples of scatter plots for multi-unit pairs with similar and different frequency tuning, and for distant and relatively close-by multi-unit clusters. Although shuffling changes the distribution of responses, separability of the 11 vocalizations does not change substantially in any of the cases. This is true across all recording sets and was shown here for 4 examples.\\ Correlated trial-variability due to interactions between neurons in the ICC does exist but does not alter separability of the multi-units' spike rates to vocalizations, although at a single neuron level this might be different.
\section{Discussion}
In this work, we analyzed similarity of multi-unit responses to a set of 11 vocalizations across the mammalian central inferior colliculus. Our findings are based on a large set of multi-unit clusters \mbox{(N~\({=}1,152\)),} of which the best frequencies span a range between \mbox{0.5-45~kHz.} The studied vocalizations are a representative set of behaviorally relevant stimuli \citep{BerrymanGuineaPigBehav1976}.\\
Neurons with similar frequency tuning (1/3 octave) have significantly higher correlation values than differently tuned neurons (Fig.\ref{fig:TonoIso}) though differences were minor and correlation varied across multi-unit cluster. Regions of similar frequency tuning, the isofrequency laminae have been suggested earlier to be functional processing units \citep{LangnerLaminar1997}. We investigated response similarity separately for these two groups.

\subsection{Spatial Dependance}
In contrast to previous studies \citep{ChenNeighborSame2012,Eggermont1987,Eggermont2006}, this work is a first systematic investigation of response similarity to vocalizations in dependance of spatial distance in the mammalian ICC.\\
Response correlations and neural correlations to artificial sound have been analyzed for single neurons in the cat ICC, for relatively short distances up to 370 \(\mu\)m \citep{ChenNeighborSame2012}, showing that nearby neurons have a higher probability of displaying similar neural preferences and responses, and suggesting a microcircuitry. 
Epping and Eggermont \citep{Eggermont1987} analyzed neural activity of 150 sorted units in the auditory midbrain of the grass frog. Using cross-correlation, they found response correlations for 60 \(\%\) of the neurons and 15 \(\%\) of the neurons displayed correlations due to neural interactions and connections. These neural correlations were restricted to pairs with distances less than 300~\(\mu\)m, whereas the response correlations were independent of the spatial distance. The authors suggested that the high amount of response correlations relative to the small amount of neural correlations indicates that in order to create the neural response, the stimulus input plays a predominant role over neural mechanisms \citep{Eggermont1987}. They attributed this to a spread projection of the stimulus input, rather than to restricted areas, which would be in line with a weak tonotopic organization in the grass frog midbrain and to the finding that neighboring neurons often display different spectral preferences.\\
In contrast to the grasshopper midbrain, the mammalian ICC has a clear tonotopic structure with neighboring neurons displaying similar spectral preferences. The correlation structure as described by Epping and Eggermont is very different for the mammalian ICC.
We showed that response similarity depends on the spatial distance between two multi-unit clusters, and decreases exponentially with increasing distance. In general, for distances above 400~\(\mu\)m (Fig.~\ref{fig:LinTono1}a,~\ref{fig:LinIso1}a,~\ref{fig:LinIso1}a,c), very little (\(\leq 0.11\)) correlation of spiking responses is present. Correlations vary across vocalizations but all display the linear dependence on spatial distance between neurons (Fig.~\ref{fig:LinVoc}). The `tooth chatter' which shows phase-locking to the stimulus envelope throughout a very large frequency range (Fig.~\ref{fig:psth}), displays the highest correlations. Responses from local field potentials have much higher correlations, and a more flat, also linear, decrease and display high correlations (\(\>0.5\)) for distances as high as 1600~\(\mu\)m (Fig.~\ref{fig:LinTono1}b,~\ref{fig:LinIso1}b,~\ref{fig:LinIso1}b,d). These large correlations lengths are due to the local field potential being long range activity. \\
The decrease could be exponentially fitted with amplitude and decay parameters ranging between \(a = 0.01-0.59\) and \(b = 0.041-0.57/\mu\)m
for spiking activity, and the ranges for similarly and differently frequency-tuned neurons overlapped in parts (Tab.~\ref{tab:fit}).
In a higher auditory processing station, in the cortex, a correlation dependance on the spatial distance has also been demonstrated. Eggermont \citep{Eggermont2006} analyzed neural groups, that reflect patched activity and were termed `clusters', in the cat primary auditory cortex with the use of cross-correlation matrices of spontaneous activity. The author found that the correlation followed an exponential decrease with distance in mm, a\(=\)0.05,~b\(=\)0.24/mm. \newpage
Our obtained values for the maximum amplitude fall within the same order, but the decay parameter \(b\) obtained in our study is at least a 17 times larger than the one found in \citep{Eggermont2006}. Eggermont found this dependence for spontaneous activity of neural groups in the cat primary auditory cortex which is also tonotopically organized. The steeper decrease in our work might be explained by the smaller size and mapping space available for the ICC compared to the primary auditory cortex, and by the different and smaller animal studied. It has been suggested that the functional organization is dynamic, and that the functional connections depend on the particular stimulus applied \citep{Eggermont1987}; thus, the use of responses to natural communications sounds instead of responses to spontaneous activity could also account for this difference in spatial decrease.\\
In this analysis, correlations for two multi-unit clusters with similar frequency tuning within 1/3 octave that are separated by at least 200~\(\mu\)m were analyzed (respectively 100 \(\mu\)m for differently tuned neurons), in order to ensure comparison of different signals. However, neural responses from a radius of more than 200~\(\mu\)m might be picked up by the recording electrode. In order to exclude this possibility, and to investigate correlation dependencies for smaller distances (\(<\)100~\(\mu\)m), single neuron recordings could be used. 
In the present analysis, we compare within 1/3 octave similarly and differently frequency-tuned multi-unit cluster. This allows one to make inferences about the organization of response similarity within the ICC insofar as neurons within an isofrequency laminae have the same best frequency within this interval of 1/3 octave, hence if the neurons differ in best frequency by more than this interval, they are likely in different isofrequency laminae. Histological stains would allow to obtain the exact location of the the multi-unit pairs within the ICC. Furthermore, knowing the positions of the neurons would allow mapping out and testing the dependence of the response similarity on the specific location within the ICC. Additionally, this study could be complemented by single neuron responses which would further reduce ambiguity of the exact position of the neural response. Another interesting and challenging investigation would be to label synaptic input in the ICC from the different ascending brainstem nuclei and measure the position of the neuron pairs relative to these main inputs in the ICC.\\
We analyzed response similarity for neurons tuned either similarly or differently to the stimulus frequency and showed a minor but significantly higher correlation for similarly frequency-tuned neurons (Fig~\ref{fig:TonoIso}) with variability. It would be informative to study similarity dependence on spatial distance with respect to neural modulation preferences, the best amplitude modulation frequency of the stimulus envelope. These neural preferences has been suggested to be arranged in a concentric gradient within isofrequency laminae \citep{LangnerPeriod1988b, Langner2002, Rees2011}. This might help to further discriminate in detail response similarity within and across isofrequency laminae. An ideal experiment to measure this would be to record simultaneously from several neurons within an isofrequency lamina using a tetrode array recordings with relatively high impedance to capture single neuron activity. Recordings would be made in response to vocalizations and in response to dynamic moving ripple sound \citep{Monty2002} which allows to compute the receptive fields and modulation preferences for the single neurons. Histological stains would inform about the actual positions of the recording sites.\\
In summary, despite the vocalizations displaying very diverse and inhomogeneous spectral contents, and despite the locally diverse inputs to the ICC, the neural representations of vocalizations exhibit gradual organization of the similarity of the neurons' responses. Multi-unit clusters with similar spiking responses are spatially spread for distances \(<\)400~\(\mu\)m, however, differences might exist for single neurons and cannot be captured at this resolution level. These findings give indications for applications in auditory midbrain prosthesis, e.g.~for the sufficient spatial separation between stimulating electrodes and contacts.
\newpage\subsection{Neural Correlations}
For our large set of 1,152 multi-unit cluster, we find that simultaneous response correlations are significantly higher than the respective non-simultaneous response correlations. Correlations due to interactions between the multi-unit cluster exist in the mammalian ICC. 
This finding is very different from a study of Epping and Eggermont \citep{Eggermont1987} in which they showed that for the grass frog midbrain only 15 \(\%\) of the units displayed neural correlations. In contrast to the grasshopper midbrain, it has been shown that the mammalian ICC has a clear tonotopic structure with neighboring neurons displaying similar spectral preferences, and a gradient of amplitude modulation frequency. We showed significant neural correlations across the ICC in response. They also show a linear decrease with spatial dependence. Thus, in the mammalian ICC neural interactions contribute significantly to shape the neural organization of vocalizations and the output of this nucleus. \\
Epping and Eggermont \citep{Eggermont1987} also found stimulus dependencies for half of the neural correlations, indicating that the functional organization is dynamic, thus the functional connections depend on the particular stimulus applied. In this work, we analyzed correlations separately for each vocalization, in order to account for such possible stimulus dependencies. The amount of neural correlations varies across vocalization stimuli (Fig.~\ref{fig:ShuffleCorrAll}), which is also true for the response correlations (Fig.~\ref{fig:TonoIso}).\\
The neural correlations though much smaller than the response correlations follow the same spatial dependence, as do the stimulus correlations. This could indicate that both effects contribute to the gradual decrease of response similarity, the gradual changing of spectrotemporal properties with spatial distance as well as neural interactions which decrease with spatial distance, since the interactions are most likely between nearby and connected neurons. The neural interactions do not contribute to a better separability between the time-averaged spike rates across similarly and differently frequency-tuned, distant or close-by neural pairs (Fig.\ref{fig:noiseCorr}), but might further shape the organization of the neural representation within the ICC.\\
Single neuron responses could be additionally recorded for this study of large spatial distances. Furthermore, single neuron recordings would allow making inferences of connectivities and quantify the amount of correlations due to neural interactions. However, limitations of the shift predictor exist even for single neurons such as the obscuring of neural interaction due to deterministic responses or temporal overlap of stimulus and neural response \citep{Eggermont1987}. Probabilistic models based on pairwise interaction, describing the weights of interactions in a network could be established \citep{Bialek2006}. \\
Anesthesia has been shown to affect neural responses \citep{SykaAnae1996}, and by fluctuating or slowly changing anesthesia levels also changes the brain state which might affect the cross-correlation strength. Thus, response similarity and particularly neural correlations might differ in awake animals, however, recordings in awake animals also bear difficulties and biases. Since anesthesia has non-negligible effects on the neural activity \citep{SykaAnae1996}, correlation values are likely to improve in awake animals.\\
In summary, it was found that multi-unit clusters in the ICC display significant neural correlations that are due to interactions of neurons. These exist for similarly and differently frequency-tuned neurons, decrease with spatial distance and differ across vocalizations. For LFPs, the neural correlations are even larger than stimulus correlations for most of the vocalizations. These findings suggests that the neural interactions shape the spiking output of this nucleus, and the neural representation of vocalizations with gradually decreasing similarity in the central IC.\\
In conclusion, we showed that despite the diverse inputs to the ICC from all ascending projections, terminating in different spatial locations in the ICC \citep{DougChapter}, and the rich spectrotemporal properties of vocalizations, the neural representation shows an organization of gradual decrease in response similarity with spatial distance, for spiking activity and for local field potentials. Thus, when comparing responses of neuronal groups to complex sound, such as vocalizations across the ICC, it is important to take their spatial separation into account and not only their frequency tuning. Our findings suggest that for multi-unit cluster in the mammalian inferior colliculus, the gradual response similarity with spatial distance to natural complex sounds is shaped by neural interactions and the gradients of neural preferences.

\section*{Acknowledgments}
We thank Thilo Rode, Tanja Hartmann, Thomas Lenarz and Hubert H. Lim for providing the neural data and vocalizations. 
This work was supported by Grant \(\#\) 01GQ0810 and \(\#\) 01GQ0811 of the Federal Ministry of Education and Research within the Bernstein Focus of Neural Technology G\"ottingen.

\bibliographystyle{jneurosci}
\bibliography{lita3}

\end{document}